%
\def\unlockat{\catcode`\@=11}
\def\lockat{\catcode`\@=12}
\unlockat
\def\d@f@ult{} \newif\ifamsfonts \newif\ifafour
%
%
%

\font\twelverm=cmr12
\font\ninerm=cmr9
\font\sixrm=cmr6
\font\fourteenbf=cmbx12 scaled\magstep1
\font\twelvebf=cmbx12
\font\ninebf=cmbx9
\font\sixbf=cmbx6
\font\fourteeni=cmmi12 scaled\magstep1      \skewchar\fourteeni='177
\font\twelvei=cmmi12                        \skewchar\twelvei='177
\font\ninei=cmmi9                           \skewchar\ninei='177
\font\sixi=cmmi6                            \skewchar\sixi='177
\font\fourteensy=cmsy10 scaled\magstep2     \skewchar\fourteensy='60
\font\twelvesy=cmsy10 scaled\magstep1       \skewchar\twelvesy='60
\font\ninesy=cmsy9                          \skewchar\ninesy='60
\font\sixsy=cmsy6                           \skewchar\sixsy='60
\font\fourteenex=cmex10 scaled\magstep2
\font\twelveex=cmex10 scaled\magstep1

\ifamsfonts
   \font\ninex=cmex9
   
   \font\sixex=cmex7 at 6pt
   
\else
   \font\ninex=cmex10 at 9pt
   
   \font\sixex=cmex10 at 6pt
   
\fi
\font\fourteensl=cmsl10 scaled\magstep2
\font\twelvesl=cmsl10 scaled\magstep1

\font\sevensl=cmsl10 at 7pt
\font\sixsl=cmsl10 at 6pt

\font\fourteenit=cmti12 scaled\magstep1
\font\twelveit=cmti12

\font\fourteentt=cmtt12 scaled\magstep1
\font\twelvett=cmtt12
\font\fourteencp=cmcsc10 scaled\magstep2
\font\twelvecp=cmcsc10 scaled\magstep1

\ifamsfonts
   
\else
   
\fi
\newfam\cpfam
\font\fourteenss=cmss12 scaled\magstep1
\font\twelvess=cmss12
\font\tenss=cmss10
\font\niness=cmss9

\font\sevenss=cmss8 at 7pt
\font\sixss=cmss8 at 6pt
\newfam\ssfam
\newfam\msafam \newfam\msbfam \newfam\eufam
\ifamsfonts
 \font\fourteenmsa=msam10 scaled\magstep2
 \font\twelvemsa=msam10 scaled\magstep1
 \font\tenmsa=msam10
 \font\ninemsa=msam9
 \font\sevenmsa=msam7
 \font\sixmsa=msam6
 \font\fourteenmsb=msbm10 scaled\magstep2
 \font\twelvemsb=msbm10 scaled\magstep1
 \font\tenmsb=msbm10
 \font\ninemsb=msbm9
 \font\sevenmsb=msbm7
 \font\sixmsb=msbm6
 \font\fourteeneu=eufm10 scaled\magstep2
 \font\twelveeu=eufm10 scaled\magstep1
 \font\teneu=eufm10
 \font\nineeu=eufm9
 
 \font\seveneu=eufm7
 \font\sixeu=eufm6
 \def\hexnumber@#1{\ifnum#1<10 \number#1\else
  \ifnum#1=10 A\else\ifnum#1=11 B\else\ifnum#1=12 C\else
  \ifnum#1=13 D\else\ifnum#1=14 E\else\ifnum#1=15 F\fi\fi\fi\fi\fi\fi\fi}
 \def\hexmsa{\hexnumber@\msafam}
 \def\hexmsb{\hexnumber@\msbfam} 
\fi
\newdimen\b@gheight             \b@gheight=12pt
\newcount\f@ntkey               \f@ntkey=0
\def\f@m{\afterassignment\samef@nt\f@ntkey=}
\def\samef@nt{\fam=\f@ntkey \the\textfont\f@ntkey\relax}
\def\rm{\f@m0 }
\def\mit{\f@m1 }
\def\cal{\f@m2 }
\def\it{\f@m\itfam}
\def\sl{\f@m\slfam}
\def\bf{\f@m\bffam}
\def\tt{\f@m\ttfam}
\def\caps{\f@m\cpfam}
\def\ssf{\f@m\ssfam}
\ifamsfonts
 \def\msa{\f@m\msafam}
 \def\msb{\f@m\msbfam} \let\bb=\msb
 \def\eu{\f@m\eufam}
\else
 \let \bb=\bf \let\eu=\bf
\fi
\def\fourteenpoint{\relax
    \textfont0=\fourteencp          \scriptfont0=\tenrm
      \scriptscriptfont0=\sevenrm
    \textfont1=\fourteeni           \scriptfont1=\teni
      \scriptscriptfont1=\seveni
    \textfont2=\fourteensy          \scriptfont2=\tensy
      \scriptscriptfont2=\sevensy
    \textfont3=\fourteenex          \scriptfont3=\twelveex
      \scriptscriptfont3=\tenex
    \textfont\itfam=\fourteenit     \scriptfont\itfam=\tenit
    \textfont\slfam=\fourteensl     \scriptfont\slfam=\tensl
      \scriptscriptfont\slfam=\sevensl
    \textfont\bffam=\fourteenbf     \scriptfont\bffam=\tenbf
      \scriptscriptfont\bffam=\sevenbf
    \textfont\ttfam=\fourteentt
    \textfont\cpfam=\fourteencp
    \textfont\ssfam=\fourteenss     \scriptfont\ssfam=\tenss
      \scriptscriptfont\ssfam=\sevenss
    \ifamsfonts
       \textfont\msafam=\fourteenmsa     \scriptfont\msafam=\tenmsa
         \scriptscriptfont\msafam=\sevenmsa
       \textfont\msbfam=\fourteenmsb     \scriptfont\msbfam=\tenmsb
         \scriptscriptfont\msbfam=\sevenmsb
       \textfont\eufam=\fourteeneu     \scriptfont\eufam=\teneu
         \scriptscriptfont\eufam=\seveneu \fi
    \samef@nt
    \b@gheight=14pt
    \setbox\strutbox=\hbox{\vrule height 0.85\b@gheight
                                depth 0.35\b@gheight width\z@ }}
\def\twelvepoint{\relax
    \textfont0=\twelverm          \scriptfont0=\ninerm
      \scriptscriptfont0=\sixrm
    \textfont1=\twelvei           \scriptfont1=\ninei
      \scriptscriptfont1=\sixi
    \textfont2=\twelvesy           \scriptfont2=\ninesy
      \scriptscriptfont2=\sixsy
    \textfont3=\twelveex          \scriptfont3=\ninex
      \scriptscriptfont3=\sixex
    \textfont\itfam=\twelveit    
    \textfont\slfam=\twelvesl    
      \scriptscriptfont\slfam=\sixsl
    \textfont\bffam=\twelvebf     \scriptfont\bffam=\ninebf
      \scriptscriptfont\bffam=\sixbf
    \textfont\ttfam=\twelvett
    \textfont\cpfam=\twelvecp
    \textfont\ssfam=\twelvess     \scriptfont\ssfam=\niness
      \scriptscriptfont\ssfam=\sixss
    \ifamsfonts
       \textfont\msafam=\twelvemsa     \scriptfont\msafam=\ninemsa
         \scriptscriptfont\msafam=\sixmsa
       \textfont\msbfam=\twelvemsb     \scriptfont\msbfam=\ninemsb
         \scriptscriptfont\msbfam=\sixmsb
       \textfont\eufam=\twelveeu     \scriptfont\eufam=\nineeu
         \scriptscriptfont\eufam=\sixeu \fi
    \samef@nt
    \b@gheight=12pt
    \setbox\strutbox=\hbox{\vrule height 0.85\b@gheight
                                depth 0.35\b@gheight width\z@ }}
\twelvepoint
%
%
\baselineskip = 15pt plus 0.2pt minus 0.1pt 
\lineskip = 1.5pt plus 0.1pt minus 0.1pt
\lineskiplimit = 1.5pt
\parskip = 6pt plus 2pt minus 1pt
\interlinepenalty=50
\interfootnotelinepenalty=5000
\predisplaypenalty=9000
\postdisplaypenalty=500
\hfuzz=1pt
\vfuzz=0.2pt
\dimen\footins=24 truecm 
\ifafour
 \hsize=16cm \vsize=22cm
\else
 \hsize=6.5in \vsize=9in
\fi
%
%
\skip\footins=\medskipamount
\newcount\fnotenumber
\def\clearfnotenumber{\fnotenumber=0} \clearfnotenumber
\def\fnote{\global\advance\fnotenumber by1 \generatefootsymbol 
 \footnote{$^{\footsymbol}$}}
\def\fd@f#1 {\xdef\footsymbol{\mathchar"#1 }}
\def\generatefootsymbol{\iffrontpage\ifcase\fnotenumber
\or \fd@f 279 \or \fd@f 27A \or \fd@f 278 \or \fd@f 27B 
\else  \fd@f 13F \fi
\else\xdef\footsymbol{\the\fnotenumber}\fi}
%
%
\newcount\secnumber \newcount\appnumber
\def\clearappnumber{\appnumber=64} \def\clearsecnumber{\secnumber=0}
\clearsecnumber \clearappnumber
\newif\ifs@c 
\newif\ifs@cd 
\s@cdtrue 
\def\unsectioned{\s@cdfalse\let\section=\subsection}
\newskip\sectionskip         \sectionskip=\medskipamount
\newskip\headskip            \headskip=8pt plus 3pt minus 3pt
\newdimen\sectionminspace    \sectionminspace=10pc
\def\Titlestyle#1{\par\begingroup \interlinepenalty=9999
     \leftskip=0.02\hsize plus 0.23\hsize minus 0.02\hsize
     \rightskip=\leftskip \parfillskip=0pt
     \advance\baselineskip by 0.5\baselineskip
     \hyphenpenalty=9000 \exhyphenpenalty=9000
     \tolerance=9999 \pretolerance=9000
     \spaceskip=0.333em \xspaceskip=0.5em
     \fourteenpoint
  \noindent #1\par\endgroup }
\def\titlestyle#1{\par\begingroup \interlinepenalty=9999
     \leftskip=0.02\hsize plus 0.23\hsize minus 0.02\hsize
     \rightskip=\leftskip \parfillskip=0pt
     \hyphenpenalty=9000 \exhyphenpenalty=9000
     \tolerance=9999 \pretolerance=9000
     \spaceskip=0.333em \xspaceskip=0.5em
     \fourteenpoint
   \noindent #1\par\endgroup }
\def\spacecheck#1{\dimen@=\pagegoal\advance\dimen@ by -\pagetotal
   \ifdim\dimen@<#1 \ifdim\dimen@>0pt \vfil\break \fi\fi}
\def\section#1{\cleareqnumber \s@ctrue \global\advance\secnumber by1
   \par \ifnum\the\lastpenalty=30000\else
   \penalty-200\vskip\sectionskip \spacecheck\sectionminspace\fi
   \noindent {\caps\enspace\S\the\secnumber\quad #1}\par
   \nobreak\vskip\headskip \penalty 30000 }
\def\undertext#1{\vtop{\hbox{#1}\kern 1pt \hrule}}
\def\subsection#1{\par
   \ifnum\the\lastpenalty=30000\else \penalty-100\smallskip
   \spacecheck\sectionminspace\fi
   \noindent\undertext{#1}\enspace \vadjust{\penalty5000}}

\def\appendix#1{\cleareqnumber \s@cfalse \global\advance\appnumber by1
   \par \ifnum\the\lastpenalty=30000\else
   \penalty-200\vskip\sectionskip \spacecheck\sectionminspace\fi
   \noindent {\caps\enspace Appendix \char\the\appnumber\quad #1}\par
   \nobreak\vskip\headskip \penalty 30000 }
\def\ack{\par\penalty-100\medskip \spacecheck\sectionminspace
   \line{\fourteencp\hfil ACKNOWLEDGEMENTS\hfil}%
\nobreak\vskip\headskip }
\def\refs{\begingroup \par\penalty-100\medskip \spacecheck\sectionminspace
   \line{\fourteencp\hfil REFERENCES\hfil}%
\nobreak\vskip\headskip \frenchspacing }
\def\endrefs{\par\endgroup}
%
%
\newif\iffrontpage \frontpagefalse
\headline={\hfil}
\footline={\iffrontpage\hfil\else \hss\twelverm
-- \folio\ --\hss \fi }
%
%
\newskip\frontpageskip \frontpageskip=12pt plus .5fil minus 2pt
\def\titlepage{\global\frontpagetrue\hrule height\z@ \relax
               \pubblock\relax }
\def\endtitlepage{\vfil\break\clearfnotenumber\frontpagefalse}
\def\title#1{\vskip\frontpageskip\Titlestyle{\caps #1}\vskip3\headskip}
\def\author#1{\vskip.5\frontpageskip\titlestyle{\caps #1}\nobreak}
\def\and{\par\kern 5pt \centerline{\sl and}}
\def\andauthor{\vskip.5\frontpageskip\centerline{and}\author}

\def\address#1{\par\kern 5pt\titlestyle{\it #1}}
\def\andaddress{\par\kern 5pt \centerline{\sl and} \address}

\def\abstract#1{\par\dimen@=\prevdepth \hrule height\z@ \prevdepth=\dimen@
   \vskip\frontpageskip\spacecheck\sectionminspace
   \centerline{\fourteencp ABSTRACT}\vskip\headskip
   {\noindent #1}}

\def\email#1{\fnote{\tentt e-mail: #1\hfill}}
\def\newaddress#1{\fnote{\tenrm #1\hfill}}
\let\thanks=\newaddress
%
%

%

%
\def\QMW{\address{%
   Department of Physics, Queen Mary and Westfield College\break
   Mile End Road, London E1 4NS, UK}}
\def\MONTP{\address{%
    Laboratoire de Physique Math\'ematique\break
    Universit\'e de Montpellier II, Place Eug\`ene Bataillon\break
    34095 Montpellier, CEDEX 5, FRANCE}}
%
%
\newcount\refnumber \def\clearrefnumber{\refnumber=0}  \clearrefnumber
\newwrite\R@fs                              
\immediate\openout\R@fs=\jobname.refs 
\def\closerefs{\immediate\closeout\R@fs} 
\def\refsout{\closerefs\refs
\unlockat
\input\jobname.refs
\lockat
\endrefs}
\def\refitem#1{\item{{\bf #1}}}
\def\ifundefined#1{\expandafter\ifx\csname#1\endcsname\relax}
\def\[#1]{\ifundefined{#1R@FNO}%
\global\advance\refnumber by1%
\expandafter\xdef\csname#1R@FNO\endcsname{[\the\refnumber]}%
\immediate\write\R@fs{\noexpand\refitem{\csname#1R@FNO\endcsname}%
\noexpand\csname#1R@F\endcsname}\fi{\bf \csname#1R@FNO\endcsname}}
\def\refdef[#1]#2{\expandafter\gdef\csname#1R@F\endcsname{{#2}}}
%
%
\newcount\eqnumber \def\cleareqnumber{\eqnumber=0}
\newif\ifal@gn \al@gnfalse  
\def\veqnalign#1{\al@gntrue \vbox{\eqalignno{#1}} \al@gnfalse}
\def\eqnalign#1{\al@gntrue \eqalignno{#1} \al@gnfalse}
\def\(#1){\relax%
\ifundefined{#1@Q}
 \global\advance\eqnumber by1
 \ifs@cd
  \ifs@c
   \expandafter\xdef\csname#1@Q\endcsname{{%
\noexpand\rm(\the\secnumber .\the\eqnumber)}}
  \else
   \expandafter\xdef\csname#1@Q\endcsname{{%
\noexpand\rm(\char\the\appnumber .\the\eqnumber)}}
  \fi
 \else
  \expandafter\xdef\csname#1@Q\endcsname{{\noexpand\rm(\the\eqnumber)}}
 \fi
 \ifal@gn
    & \csname#1@Q\endcsname
 \else
    \eqno \csname#1@Q\endcsname
 \fi
\else%
\csname#1@Q\endcsname\fi\global\let\@Q=\relax}
%
%
\newif\ifm@thstyle \m@thstylefalse
\def\mathstyle{\m@thstyletrue}
\def\proclaim#1#2\par{\smallbreak\begingroup
\advance\baselineskip by -0.25\baselineskip%
\advance\belowdisplayskip by -0.35\belowdisplayskip%
\advance\abovedisplayskip by -0.35\abovedisplayskip%
    \noindent{\caps#1.\enspace}{#2}\par\endgroup%
\smallbreak}
\def\m@kem@th<#1>#2#3{%
\ifm@thstyle \global\advance\eqnumber by1
 \ifs@cd
  \ifs@c
   \expandafter\xdef\csname#1\endcsname{{%
\noexpand #2\ \the\secnumber .\the\eqnumber}}
  \else
   \expandafter\xdef\csname#1\endcsname{{%
\noexpand #2\ \char\the\appnumber .\the\eqnumber}}
  \fi
 \else
  \expandafter\xdef\csname#1\endcsname{{\noexpand #2\ \the\eqnumber}}
 \fi
 \proclaim{\csname#1\endcsname}{#3}
\else
 \proclaim{#2}{#3}
\fi}
\def\Thm<#1>#2{\m@kem@th<#1M@TH>{Theorem}{\sl#2}}
\def\Prop<#1>#2{\m@kem@th<#1M@TH>{Proposition}{\sl#2}}
\def\Def<#1>#2{\m@kem@th<#1M@TH>{Definition}{\rm#2}}
\def\Lem<#1>#2{\m@kem@th<#1M@TH>{Lemma}{\sl#2}}
\def\Cor<#1>#2{\m@kem@th<#1M@TH>{Corollary}{\sl#2}}
\def\Conj<#1>#2{\m@kem@th<#1M@TH>{Conjecture}{\sl#2}}
\def\Rmk<#1>#2{\m@kem@th<#1M@TH>{Remark}{\rm#2}}
\def\Exm<#1>#2{\m@kem@th<#1M@TH>{Example}{\rm#2}}
\def\Qry<#1>#2{\m@kem@th<#1M@TH>{Query}{\it#2}}
%
%

%
\def\<#1>{\csname#1M@TH\endcsname}
%
%
\def\ref#1{{\bf [#1]}}
\def\ie{{\it i.e.\/}}
%
%

\def\lapprox{\hbox{\lower3pt\hbox{$\buildrel<\over\sim$}}}
\def\gapprox{\hbox{\lower3pt\hbox{$\buildrel<\over\sim$}}}
\def\quotient#1#2{#1/\lower0pt\hbox{${#2}$}}
\def\fr#1/#2{\mathord{\hbox{${#1}\over{#2}$}}}
\ifamsfonts
 \mathchardef\empty="0\hexmsb3F 
 \mathchardef\lsemidir="2\hexmsb6E 
 \mathchardef\rsemidir="2\hexmsb6F 
\else
 \let\empty=\emptyset
 \def\lsemidir{\mathbin{\hbox{\hskip2pt\vrule height 5.7pt depth -.3pt
    width .25pt\hskip-2pt$\times$}}}
 \def\rsemidir{\mathbin{\hbox{$\times$\hskip-2pt\vrule height 5.7pt
    depth -.3pt width .25pt\hskip2pt}}}
\fi
%
%

%
%
\def\reals{\mathord{\bb R}} 
%
%
\def\underrightarrow#1{\vtop{\ialign{##\crcr
      $\hfil\displaystyle{#1}\hfil$\crcr
      \noalign{\kern-\p@\nointerlineskip}
      \rightarrowfill\crcr}}} 
\def\underleftarrow#1{\vtop{\ialign{##\crcr
      $\hfil\displaystyle{#1}\hfil$\crcr
      \noalign{\kern-\p@\nointerlineskip}
      \leftarrowfill\crcr}}}  

%
%
%
%

\def\NPB#1#2#3{{\sl Nucl. Phys.} {\bf B#1} (#2) #3}

\def\CMP#1#2#3{{\sl Comm. Math. Phys.} {\bf #1} (#2) #3}
\def\PRD#1#2#3{{\sl Phys. Rev.} {\bf D#1} (#2) #3}

\def\PLB#1#2#3{{\sl Phys. Lett.} {\bf #1B} (#2) #3}

\def\PR#1#2#3{{\sl Phys. Reports} {\bf #1} (#2) #3}

\def\FAaIA#1#2#3{{\sl Functional Analysis and Its Application} {\bf #1} (#2) #3}

\def\AIHP#1#2#3{{\sl Ann. Inst. Henri Poincar\'e} {\bf #1} (#2) #3}

\def\JPA#1#2#3{{\sl J. Physics} {\bf A#1} (#2) #3}

\def\MPLA#1#2#3{{\sl Mod. Phys. Lett.} {\bf A#1} (#2) #3}

\lockat


%
%
\def\W{\mathord{\ssf W}}

\def\fr#1/#2{\mathord{\hbox{${#1}\over{#2}$}}}

\def\ket|#1>{\mathord{\vert{#1}\rangle}}

\def\ope#1#2{{{#2}\over{\ifnum#1=1 {z-w} \else {(z-w)^{#1}}\fi}}}

\def\corr<#1>{\mathord{\langle #1 \rangle}}

\def\Im{\mathop{\rm Im}}
\def\Re{\mathop{\rm Re}}
\def\KdV{{\ssf KdV}}
\def\x{{\bf x}}
\def\p{{\bf p}}
\def\z{{\bf z}}
\def\P{{\bf P}}
\def\v{{\bf v}}
\def\u{{\bf u}}
\def\Pau{{\bf S}}
\def\complex{{\bb C}}
\def\compro{{\bb C}{\bb P}}

\font\tiny=cmr10
%
%
\refdef[Sniatycki]{J. Sniatycki, \AIHP{20}{1974}{365}.}
\refdef[Woodhouse]{N.M.J. Woodhouse, {\sl Geometric Quantization},
Oxford Science Publications, Oxford 1991.}
\refdef[Polyakov]{A.M. Polyakov, \NPB{268}{1986}{406}.}
\refdef[Pisarski]{R.D. Pisarski, \PRD{34}{1986}{670}.}
\refdef[PolyakovII]{A.M. Polyakov, \MPLA{3}{1988}{325}.}
\refdef[RamosRocaI]{E. Ramos and J. Roca, \NPB{452}{1995}{705},
({\tt hep-th/9504071}).}
\refdef[RamosRocaII]{E. Ramos and J. Roca, \NPB{436}{1995}{529},
({\tt hep-th/9408019}).}
\refdef[Radul]{A.O.Radul,\FAaIA{25}{1991}{25}.}
\refdef[Bacry]{H. Bacry, \CMP{5}{1967}{97}.}
\refdef[Gomis]{C. Batlle, J. Gomis, J.M. Pons and N. Roman-Roy,
\JPA{21}{1988}{2693};
X. Gr\`acia and J.M. Pons, Gauge transformations
for higher order lagrangians, preprint UB-ECM-PF-95-15,
({\tt hep-th/95090954}).}
\refdef[Plyushchay]{M.S. Plyushchay, \MPLA{4}{1989}{837}.}
\refdef[PlyushchayII]{M.S. Plyushchay, \NPB{389}{1993}{181}.}
\refdef[Penrose]{R. Penrose and W. Rindler, {\sl Spinors and space-time},
Cambridge University Press, Cambridge 1984.}
\refdef[PenroseII]{R. Penrose and M. MacCullan, \PR{C6}{1972}{247}.}
\refdef[Wiegmann]{P.B. Wiegmann, \NPB{323}{1989}{311}.}
\refdef[Forte]{S. Forte, \PLB{253}{1991}{403}.}
\refdef[Weinstein]{A. Weinstein, {\sl Lectures on symplectic manifolds}.
CBMS regional conference series, Vol. 29. Amer. Math. Soc., Providence,
Rhode Island 1977.}
\refdef[Coherent]{S. Iso, C. Itoi and H. Mukaida, \NPB{346}{1990}{293}.}

%

%
\overfullrule=0pt
\def\pubblock{ \line{\hfil\rm PM/95-52}
               \line{\hfil\rm QMW--PH--95--49}
               \line{\hfil\tt hep-th/9601035}
               \line{\hfil\rm December 1995}}
\titlepage
\title{On the $\W$-geometrical origins of massless field equations
and gauge invariance}

\author{Eduardo Ramos\email{ramos@lpm.univ-montp2.fr}}
\MONTP
\andauthor{Jaume Roca\email{J.Roca@qmw.ac.uk}\newaddress{Address after
1st January 1996: Dept. d'Estructura i Constituents de la Mat\`eria,
U. Barcelona, Diagonal 647, E-08028 Barcelona, Catalonia, Spain.}}
\QMW
\vskip 5truecm
\eject
\abstract{
We show how to obtain all covariant field equations for massless particles of
arbitrary integer, or half-integer, helicity in four dimensions from the
quantization of the rigid particle, whose action is given by the integrated
extrinsic curvature of its worldline, {\ie} $S=\alpha\int ds \kappa$. This
geometrical particle system possesses one extra gauge invariance besides
reparametrizations, and the full gauge algebra has been previously identified
as classical $\W_3$. The key observation is that the
covariantly reduced phase space of this model can be naturally identified
with the spinor and twistor descriptions of the covariant phase spaces
associated with massless particles of helicity $s=\alpha$. Then, standard
quantization techniques require $\alpha$ to be quantized and show how the
associated Hilbert spaces are solution spaces of the standard relativistic
massless wave equations with $s=\alpha$. Therefore, providing us with a simple
particle model for Weyl fermions ($\alpha=1/2$), Maxwell fields ($\alpha=1$),
and higher spin fields. Moreover, one can go a little further and in the
Maxwell case show that, after a suitable redefinition of constraints, the
standard Dirac quantization procedure for first-class constraints leads to a
wave-function which can be identified with the gauge potential $A_\mu$. Gauge
symmetry appears in the formalism as a consequence of the invariance under
$\W_3$-morphisms, that is, exclusively in terms of the extrinsic
geometry of paths in Minkowski space.
When all gauge freedom is fixed one naturally obtains
the standard Lorenz gauge condition on $A_{\mu}$, and Maxwell equations in
that gauge. This construction has a direct generalization to
arbitrary integer values of $\alpha$, and we comment on the physically
interesting case of linearized Einstein gravity ($\alpha =2$).
}
\endtitlepage
\section {Introduction}

It was not until recently that geometrical particle models, other than
the one associated with the worldline length, came to attract some
attention from the physics comunity. And even then, they were only
considered \[Pisarski] as toy models for rigid strings \[Polyakov], or
as the simplest non-trivial examples to test
the formalism of singular higher-order derivative theories
\[Gomis].
Nevertheless, it was soon
realized, mainly due to the pionnering work of Plyushchay, that these systems
were interesting in their own right. It was shown in \[Plyushchay] how a
noncovariant quantization of the rigid particle in Minkowski space, whose
action is given by
$$S=\alpha\int ds\ \kappa, \(rigidaction)$$
where $\alpha$ is a dimensionless coupling constant, and the extrinsic
curvature $\kappa$ is given by
$$\kappa =\left|g_{\mu\nu} {d^2 x^{\mu}\over ds^2}{d^2 x^{\nu}\over
ds^2} \right|,\(defkappa)$$
provides us with a potential particle candidate for the description of
photons and other higher order spin fields. Notwithstanding the interest
of these results they fell short of proving this connection, {\ie} it
was not possible to obtain directly from this approach the
associated Poincar\'e covariant fields theories.

After these first steps a plethora of new results have emerged in this
field. Among them one should stress the ones related to
Fermi-Bose transmutation in three dimensions in the presence of a
Chern-Simons field. Polyakov \[PolyakovII]
was the first to point out that the presence
of a torsion term in the effective action for the Wilson loops was 
responsible for the appearence of Dirac fermions in an otherwise 
apparently bosonic theory.
In particular, it was again Plyushchay \[PlyushchayII] who realized
that the Dirac equation naturally appears in a fully Lorentz covariant
canonical quantization of a particle model with an extra torsion term in
$2+1$-dimensions (although by then there were already alternative proofs
of Polyakov's results based on coherent state path integrals \[Coherent]).
More recently, it was shown by the authors \[RamosRocaI] that the
extended gauge invariance present in some
of these  geometrical particle models could be naturally identified with
the classical limit
of $\W_n$-algebras. Moreover, it was shown how
the corresponding gauge transformations could be understood geometrically
through the (generalized) Gauss map of their particle trajectories.
Therefore providing a natural geometrical and dynamical framework
for $\W$-symmetry.
In particular, for the rigid particle model \(rigidaction)
it was proven \[RamosRocaII] that its gauge symmetry algebra
could be identified with the classical limit of Zamolodchikov's 
$\W_3$-algebra.
Interestingly enough the proof was based on a previously unsuspected
connection with integrable systems of the $\KdV$-type.

The purpose of this paper is to fill the gap in the results of
\[Plyushchay] and quantize the rigid particle in a fully covariant
manner. On our way we will encounter some beautiful geometric
structures associated with the reduced phase space of the system
under its $\W_3$ invariance. We will show how the space of gauge
invariant functions in phase space coincides with the one naturally
associated with massless particles of helicity $\alpha$
(the coupling constant) obtained through the coadjoint orbit
method applied to the Poincar\'e group \[Woodhouse]. It is then
a standard exercise in quantization to show that
$\alpha$ is quantized and can take only integer or
half integer values. Moreover, the Hilbert spaces
in spinor or twistor (polarizations) coordinates are easily constructed
and they are found to be, respectively, the solution spaces
for the standard relativistic wave equations in spinor or twistor
representation \[Penrose] with
helicity $\alpha$. In particular, $\alpha = 1/2,1,2$
correspond to the physically relevant cases of Weyl, Maxwell and
linearized Einstein gravity field equations.

We will also explicitly show how to recover in the case $\alpha=1$ the
standard gauge potential ($A_{\mu}$) description of Maxwell equations.
This is achieved by recasting the first-class constraints
of the model in spinor formalism, and quantizing them {\it \`a la}
Dirac. It will then be possible to understand the standard $U(1)$ gauge
symmetry of the wave function as a consequence of the
$\W_3$ gauge structure of the model, or equivalently in terms of
the extrinsic geometry of paths in Minkowski space.
The Lorenz gauge
condition and Maxwell equations for $A_{\mu}$ will naturally appear
from Dirac's prescription by imposing the first-class
constraints as operator constraints in the wave function.
This construction has a direct generalization to arbitrary integer
$\alpha$. We finish by comenting on its
geometrical consequences in the case of linearized Einstein gravity,
{\ie}, $\alpha=2$.

In order to be reasonably self-contained we will introduce
the necessary geometrical concepts as they are needed, and
will provide
the reader with the minimally required knowledge about the rigid
particle and its $\W_3$ gauge invariance.

\section{The rigid particle}

Let us briefly review some known results concerning the rigid
particle model.
Consider a curve $\gamma$ describing the trajectory of a particle
in Minkowski space
$$
\eqnalign{
\gamma:\;&[t_0,t_1]\longrightarrow {\bb M}^4\cr
&\quad t\quad\mapsto\quad \x(t),
\(gamma)}
$$
where we use the metric $g={\rm diag}(+---)$.
We will not require the normalized tangent vector $\v_1=d\x/ds$ to be
time-like but rather space-like, $\v_1^2=-1$. This may seem surprising
at first but it can be shown \[Plyushchay] that the constraints placed by the
dynamics of the rigid particle are only consistent in this regime. The
reader may think at this point that this space-like character of the paths
will render the theory acausal. That this is not the case can only be 
understood in terms of the extra gauge invariance of the system. It was shown 
in \[Plyushchay] how physical (gauge invariant) quantities follow a perfectly
consistent standard relativistic motion. We will try to give an intuitive
geometric picture of this fact at the end of this section when the reader
has already become acquainted with the inner workings of the model.

The extrisic curvature $\kappa$ is defined as the modulus of
$d\v_1/ds$:
$$
{d\v_1\over ds}=\kappa\;\v_2,
\(Frenet)
$$
where $\v_2$ is orthogonal to $\v_1$ and,  for later consistency
with the dynamics, we assume it to be also space-like $\v_2^2=-1$.

The coordinate expressions of $\v_1$, $\v_2$ and $\kappa$ are given by
$$
\v_1={\dot\x\over\sqrt{-\dot\x^2}},\quad
\v_2={\ddot\x_\perp\over\sqrt{-\ddot\x^2_\perp}},\quad
\kappa=-{\sqrt{-\ddot\x^2_\perp}\over\dot\x^2}>0,
$$
where $\dot \x={d\x/dt}$ and $\ddot\x_\perp=\ddot\x
-\dot\x(\ddot\x\dot\x)/\dot\x^2$.

Now the rigid particle action is defined as the integrated curvature
over the worldline:
$$
S[\x]=\alpha\int ds\;\kappa=\alpha\int dt \sqrt{\ddot\x^2_\perp \over
\dot\x^2}.
\(action)
$$

This is a higher derivative model and we expect its phase space
to be larger than the standard cotangent bundle over Minkowski space,
which is described solely by the position and total momentum coordinates
$(\x,\P)$. In the case at hand the phase space contains
an additional canonical pair $(\dot\x,\p)$.
This can be understood by noting that an arbitrary infinitesimal
variation of the action
$$
\delta S = - \int^{t_1}_{t_0} dt \; \dot \P \delta \x
\; +  \P\delta \x\Bigl\vert^{t_1}_{t_0} + \p\delta\dot\x
\Bigl\vert^{t_1}_{t_0},
\(deltaS)
$$
where
$$
\p ={\partial L\over\partial\ddot\x},\quad\quad
\P ={\partial L\over\partial\dot\x}-\dot\p,
$$
requires not only the equations of motion $\dot\P=0$ to be satisfied
with fixed endpoints, but also $\dot\x$ should be kept fixed at the
endpoints.

Thus, phase space is described by coordinates $(\x,\P,\dot\x,\p)$
and is endowed with the canonical symplectic form
$$
\Omega=d\x\wedge d\P+d\dot\x\wedge d\p.
$$

The lagrangian expression of the momenta is given by
$$
\eqnalign{
\p &=-{\alpha\over\sqrt{-\dot\x^2}}\v_2,
\cr
\P &=\alpha\left({d\v_2\over ds}+\kappa\v_1\right).
\(momenta)}$$

{}From these expressions and the Frenet equation \(Frenet) it is easy to
show that the $\v_1$, $\v_2$ and $\P$, form a triad of mutually
orthogonal vectors. Moreover, consistency of the equations of motion
$d\P/ds=0$  with the condition $\P\v_2=0$ implies that $\P$ has to be a
light-like vector. Indeed,
$$
0={d\P\over ds}\v_2+\P{d\v_2\over ds}=\P\left({\P\over\alpha}
-\kappa\v_1\right)={1\over\alpha}\P^2.
$$

All these conditions provide a set of constraints in phase space,
$\phi_i\approx0$, with
$$
\eqalign{
\phi_1=\p\dot \x,\quad\quad
&\phi_2={1\over2}\left({\dot\x^2\p^2\over\alpha}-\alpha\right),
\cr
\phi_3=\P\dot \x,\quad\quad
&\phi_4=\P\p,
\cr
&\phi_5=\P^2,}
\(constraints)
$$
which turn out to be first-class. It is customary to denote the constraints
coming from the definition of the mometum associated with the
highest-order time derivative of $\x$ as primary. In our case
$\phi_1$ and $\phi_2$ are the primary constraints and
as such they will play an essential role in the
reduction process.

First-class constraints generate gauge transformations and the model
is certainly invariant under reparametrizations of the worldline. There
is, however, an additional gauge symmetry, very peculiar of this
model which renders the position of the path as an unphysical (not gauge
invariant) quantity. This extra gauge invariance can be given
a simple geometrical interpretation as follows \[RamosRocaI]. From the
curve $\gamma$ parametrized by $\x(t)$ we can construct a
new curve $\Gamma$ (the Gauss map) which is given by the normalized
tangent vector $\v_1(t)$. Then the action \(action) is nothing more
than the arc-length of this new curve:
$$
S=\alpha\int dt\sqrt{-\dot\v^2_1}.
$$
It is clear that there are many different curves sharing the same
Gauss map and this can be seen to define the gauge orbits of this extra
symmetry. The fact that spacetime trajectories are not physical
explains why there should be no a priori inconsistency between
the space-like character of the curves and perfectly causal propagation.
Indeed, we have explicitly shown how the momentum $\P$ of the particle
(which is gauge invariant) has a perfectly well-behaved light-like
character.

It was proven in \[RamosRocaII] that the full gauge symmetry algebra of 
the rigid particle is precisely ${\W}_3$. This is most easily done by 
realising that the equations of motion\fnote{\tiny The invariance of the 
action can be equally checked by purely algebraic methods.}
can be written in terms of the Boussinesq
Lax operator. Then, standard methods in integrable systems of the $\KdV$-type
show \[Radul] that its symmetry algebra is nothing but the Gel'fand-Dickey
bracket associated with $SL(3)$, or equivalently the classical limit of
Zamolodchikov's $\W_3$-algebra. Therefore establishing a direct connection
between the extrinsic geometry of paths in Minkowski, or Euclidean, space-time
and $\W_3$.

\section{The covariantly reduced phase space of the rigid particle }

We will now perform the covariant reduction of our phase space.
We will proceed step by step because on our way some natural
mathematical structures, that will be useful in what
follows, will surface.

Let us now introduce the following complex coordinates in phase space:
$$\z=\dot\x +i {\dot\x^2\over\alpha}\p,\(complexcoordinate)$$
and $\bar\z$ its complex conjugate. Notice now that the two primary
first-class constraints define a quadric on
$\complex^4$, {\ie}, $\z\cdot\z =0$.
We can now pass to study the action of these two primary constraints on the
quadric, or equivalently, their gauge orbits. A simple computation
yields
$$\{\z,\phi_1\}=\z,\quad{\rm and}\quad \{\z,\phi_2\}=-i\z.\(gaugeorbits)$$
Which implies that the flows generated by these two constraints correspond to
multiplication by an arbitrary complex number.
Therefore if we quotient the phase space with respect to
these gauge orbits the reduced phase space (with respect
the two primary constraints)
is the standard cotangent bundle over Minkowski space-time
times a quadric in $\compro^3$.

This quadric in $\compro^3$ has a natural
geometric interpretation in terms of the Grassmann manifold of
space-like two-planes in Minkowski space, which will be denoted by
$G^M_{(2,4)}$.
This geometrical identification comes as follows:
any space-like two-plane in ${\bb M}^4$ is completely determined by
two four-vectors
$\u_1$ and $\u_2$ which are linearly independent and mutually
orthogonal, {\ie}, $\u_1\cdot\u_2=0$. Moreover, without loss of
generality one can choose that $|\u_1|=|\u_2|$.
Then, if we define $\z =\u_1 +i\u_2$ it follows that $\z\cdot\z=0$.
It is obvious that if we multiply $\z$ by an arbitrary
complex number we are still describing the same plane, because the
result
on the $\u$'s will simply amount to a combined dilatation and rotation.
Equivalently, any $\z$ belonging to this quadric in $\compro^3$ describes
uniquely a space-like plane by choosing $\u_1=\Re\z$ and $\u_2=\Im\z$.
Notice that the cases in which one or both vectors are time-like
or null are directly ruled out.

It is now straightforward to check that the choice of standard
inhomogeneous coordinates in the grassmannian, {\it i.e.},
$$\z = (1, z^1, z^2,z^3),\(homogeneous)$$
corresponds to the non-covariant gauge-fixing conditions
$\dot x^0=1$ and $p_0=0$, which were used by Plyushchay in \[Plyushchay].

Anyhow, we can now proceed in a manifestly covariant manner to
compute the symplectic form induced on the grassmannian $G^M_{(2,4)}$
(for the time being we will ignore the term $d\x\wedge d\P$ because it will be
irrelevant for this part of the discussion).
Notice that
$$\eqnalign{d\dot\x &= {1\over 2}(d\z +d\bar\z)\(difuno)\cr
    d\p &=-{i\alpha\over \z\bar\z}(d\z -d\bar\z) +
  {i\alpha\over (\z\bar\z)^2}(\bar\z d\z +\z d\bar\z)(\z -\bar\z).\(difdos)\cr}
$$
And from here one obtains 
$$d\dot\x\wedge d\p = {i\alpha\over (\z\bar\z)^2}
\left( (\z\bar\z) d\z\wedge d\bar\z - (\bar\z d\z)\wedge (\z d\bar\z)\right),
$$
where we have used in a crucial manner the fact that
that $\z d\z$ is zero in $G^M_{(2,4)}$.
This symplectic form on the grassmannian is precisely the
one naturally induced from its embedding
in $\compro^3$ with the standard Fubini-Study symplectic
form\fnote{\tiny
The reader familiar with the Dirac bracket formalism may be suspicious
that we have been oblivious to it. This is not the case, as the
Dirac bracket of functions defined on the reduced space certainly
coincides with the action of the reduced symplectic form on their
associated hamiltonian vector fields. If one chooses to
compute Poisson brackets on the constrained surface while still using
the $\z$ and $\bar \z$ coordinates, the gradients of the associated
functions turn out to be ill-defined.
By imposing on them that their hamiltonian
vector fields be tangent to the constrained manifold, one easily recovers
the standard Dirac bracket formalism \[Sniatycki].}.

The key observation which will pave our way for the study of the
reduced phase space is
that points in $G^M_{(2,4)}$ can be understood as complex null lines
passing through the origin on $\complex^4$ equipped with a minkowskian
metric. This suggests that the appropriate
Lorentz invariant formalism is supplied by the standard
spinor representation of these null lines.
Indeed, the spinor formalism \[Penrose] will turn out to be a
powerful tool in what follows.

Given an arbitrary complex four-vector ${\bf y}$ it can be rewritten
in spinor coordinates as follows:
$$
(y^{A\dot U})={1\over\sqrt 2}\left(
\matrix{    y^0+y^3   &   y^1+iy^2 \cr
            y^1-iy^2  &   y^0-y^3     }\right),
$$
so that $\det(y^{A\dot U})={1\over2}g_{\mu\nu}y^\mu y^\nu$.

Because of the two-to-one local isomorphism between $SL(2,\complex)$ and
the identity component of the Lorentz group, one such Lorentz
transformation on ${\bf y}$ is equivalently represented by the action
of an $SL(2,\complex)$ matrix acting on the undotted indices and its
complex conjugate matrix on the dotted ones.
Raising and lowering of indices is mimicked in spinor language by
contraction with the invariant antisymmetric tensors $(\epsilon^{AB})=
(\epsilon_{AB})$, with $\epsilon^{01}=+1$, and analogous expressions for
the dotted indices.

Using the antisymmetry of the invariant tensor $\epsilon$ one
finds for any (commuting) spinors $\alpha$ and $\beta$ that
$\alpha_A\beta^A=
-\alpha^A\beta_A$ and $\alpha_A\alpha^A=0$, and similarly for dotted
spinors.

Coming back to the vector $\z$ defined in \(complexcoordinate), in spinor
coordinates $$z^{\mu}\longrightarrow z^{A\dot U},\(spindes)$$
and the fact that $\z$ is null, directly implies that
$$ z^{A\dot U}=\xi^A\bar\eta^{\dot U}.\(nullcomplex)$$
Here $\xi$ and $\bar\eta$ are complex spinors, which
means eight real degrees of freedom. However, $\z$ is insensitive to
a rescaling $\xi\rightarrow a\xi$, $\bar\eta\rightarrow\bar\eta/a$,
with $a\in\complex$. Moreover, because of our freedom to rescale $\z$ itself
the spinors $\xi$ and $\bar\eta$ are both defined only up to
an arbitrary complex factor. This again reduces the number of degrees of
freedom down to four, in complete agreement with standard
hamiltonian counting.

It is then a direct computation to check that
the symplectic form in spinor coordinates is given by
$$ -i\alpha {d\eta_A\wedge d\xi^A\over \xi^C\eta_C}
+i\alpha {\eta_A\xi^B d\eta_B\wedge d\xi^A\over (\xi^C\eta_C)^2}
+ c.c. \(omegaspin)$$

Notice now that the three remaining constraints ($\phi_3,\phi_4,\phi_5$)
that we have been ignoring so far, can be neatly written in spinor form.
On the one hand the fact that $\P^2=0$ implies that $\P$ is a real null
vector and hence can be written as
$$P^{\mu} \longrightarrow P^{A\dot A} =\pm\pi^A\bar\pi^{\dot A},
\(nullmomenta)$$
where $\pi$ is completely determined up to an arbitrary phase factor,
and the plus and minus signs correspond to future or past pointing null vectors
respectively. And on the other hand
$$\P\z =0\longrightarrow (\pi_A\xi^A)(\bar\pi_{\dot A}
\bar\eta^{\dot A})=0,\()$$
so in its spinor form this constraint reduces to either
$$\pi_A\xi^A=0\quad {\rm or}\quad \bar\pi_{\dot A}\bar\eta^{\dot A}=0.
\(spinconstraints)$$

First notice that both conditions cannot be simultaneously fulfilled
because it contradicts the two primary
constraints $\phi_1$ and $\phi_2$. In fact, if both conditions were
obeyed then $\z$ would be proportional to $\P$ and therefore not
only $\z^2=0$ but also $\z\bar\z=0$, which would require $\dot\x$
to be null in clear contradiction to $\phi_2$.
Therefore the reduced phase space has four different branches that
will be denoted by $M_\alpha^{+}$, $M_{-\alpha}^{+}$, ${M}_\alpha^{-}$
and  ${M}_{-\alpha}^{-}$, where the superscript $+$ ($-$) corresponds
to future (past) pointing momentum, and the subscripts $\pm\alpha$ correspond
to different values of the helicity that will correspond, as we will show
below, to the two possible choices between the spinor constraints
\(spinconstraints).

Let us recall that the (physical) irreducible representations of
the Poincar\'e
algebra are labeled by the values of the two casimirs $\P^2 =m^2$ and the
square of the Pauli-Lubanski vector
$$S^{\mu} = {1\over 2}\epsilon^{\mu\nu\rho\sigma} P_\nu M_{\rho\sigma}.
\(PauliLub)$$
In the massless case, if one disregards the unphysical situation when
$\Pau^2 \neq 0$, it directly follows that $\Pau = s\P$ for some $s$. This
invariant is usually denoted as the helicity. We can now show that our
case will fall under this category.

First from the constraint $\P^2 =0$ it directly follows that we are dealing
with the massless case. Moreover, the ``internal'' angular momentum
matrix is given by
$$M^{\mu\nu}=\dot x^{[\mu}p^{\nu]}= -{i\alpha\over\z\bar\z}
\bar z^{[\mu}z^{\nu]}.
\()$$
And from this it follows that
$$S_\mu ={1\over 2}
\epsilon_{\mu\nu\rho\sigma}P^\nu M^{\rho\sigma}=
 -{i\alpha\over\z\bar\z}\epsilon_{\mu\nu\rho\sigma}
P^\nu\bar z^\rho z^\sigma.\(Paulirigido)$$
In spinor language the above expression reads
$$S_{A\dot A} = \pm{\alpha\over (\xi\eta)(\bar\xi\bar\eta)}
(\epsilon_{AD}\epsilon_{BC}\epsilon_{\dot A\dot C}
\epsilon_{\dot B\dot D} - \epsilon_{AC}\epsilon_{BD}\epsilon_{\dot A\dot D}
\epsilon_{\dot B\dot C}) \pi^B\bar\pi^{\dot B}
\bar\xi^{\dot C}\eta^C \xi^D\bar\eta^{\dot D}.\()
$$
Note that if we choose future pointing $\P$ and the branch of
the constraint surface given by $\pi_A\xi^A=0$ it follows that
$$S_{A\dot A} =\alpha\pi_A\bar\pi_{\dot A},\()$$
or equivalently $\Pau = s \P$ with helicity $s=\alpha$.
But if the other branch $\pi_A\eta^A$ had been chosen we would have got
a similar result, but now with a value $-\alpha$ for the helicity. The
past pointing $\P$ can be similarly worked out, thus justifying our
notational choice.

If one considers the whole Poincar\'e group and not only its connected
component,
it can be shown \[Woodhouse] that a Lorentz transformation preserving spatial
orientation, but reversing the arrow of time, maps $M^+_s$ into $M^-_{-s}$;
and maps $M^+_s$ into $M^-_s$ whenever it  reverses both space and
time orientations.
It is therefore natural to identify those subspaces and one can regard the
phase space as the union of $M^+_s$ and  $M^+_{-s}$. This can be
easily understood in our model because time reversal maps the equivalence class
of $\z$ into the one of $\bar \z$ thus interchanging the two possible branches
of our constraint \(spinconstraints).

Now we come back to our reduced phase space.
We recall that $\xi$ and $\eta$ were both defined
up to a multiplication by an arbitrary complex number. On the
constraint surface given by $\pi_A\xi^A=0$ the spinor $\xi$ must be
proportional to $\pi$, so we can remove the above freedom in $\xi$ and
$\eta$ by setting
$$
\xi^A =\pi^A\quad {\rm and}\quad \eta^A\pi_A=1, \(partialfixing)
$$
{\it i.e.}, $\pi$ and $\eta$ form a spinor basis. Notice also that because
of the phase ambiguity in $\pi$ we have an equivalent
ambiguity left in $\eta$. This corresponds to a reduction of the subspace
previously denoted by $M^{+}_{\alpha}$ if one chooses future pointing
$\P$.

One can now compute the induced symplectic form on the
submanifold defined by \(partialfixing). One readily obtains:
$$  \pi_A  dx^{A\dot A}\wedge d\bar\pi_{\dot A}
+ i\alpha  d\eta_A\wedge d\pi^A + c.c.\(formaprereducida)$$

It follows from its definition that the above form is degenerate.
This is what is to be expected in the case of first-class constraints unless
one introduces enough gauge conditions to turn all of them into second
class. This is indeed the case here. We have already solved all the
constraints $\phi_1,\ldots,\phi_5$ but have only performed the
quotient over the orbits generated by $\phi_1$ and $\phi_2$. Therefore
before continuing one should identify which vectors are in the kernel
of this form, {\it i.e.} determine the vector fields tangent to the
remaining orbits. With a little of hindsight due to the particular
structure of the constraints it is simple to check that
$$\eqnalign{
X_1 =& \pi^A\bar\pi^{\dot A}{\partial\ \over \partial
x^{A\dot A}},\(vecuno)\cr
X_2 =& \pi_A {\partial\ \over\partial \eta_A}
+i\alpha\bar\eta^{\dot A}\pi^A{\partial\ \over\partial x^{A\dot A}},
\(vecdos)\cr
X_3 =& i\eta^A{\partial\ \over\partial\eta^A}
-i\pi_A{\partial\ \over\partial\pi_A} + c.c.\cr
}$$
span the kernel of \(formaprereducida).
Notice that $X_2$ is complex so the real dimension of the space
spanned by $X_1$ and $X_2$ is
three, in full agreement with the dimension of the orbits generated by
$\phi_3$, $\phi_4$ and $\phi_5$. The vector field $X_3$ is responsible
for the phase shifts in $\pi$ and $\eta$.
With the above result in mind one could directly apply the
general reduction procedure of \[Weinstein], however things are greatly
simplified by realising that $\pi_A$ and $\omega^A$ with
$$\omega^A = \alpha\eta^A + ix^{A\dot A}\bar\pi_{\dot A}$$
are constant along the orbits generated by $X_1$ and $X_2$.
So they are natural coordinates to describe the fully
reduced phase space. From all of this it follows that
$M^{+}_{\alpha}$ can be
parametrized by the four components of the two spinors $\omega$ and
$\bar\pi$ subject to the equivalence relation
$$(\omega ,\bar\pi)\equiv ({\rm e}^{i\beta}\omega,{\rm e}^{i\beta}\bar\pi )$$
with $\beta\in\reals$  and obeying
$$
\omega^A\pi_A + \bar\omega^{\dot A}\bar\pi_{\dot A}= 2 \alpha,\(conspin)
$$
which is just the constraint $\eta^A\pi_A=1$ in \(partialfixing) written
in the $(\omega,\pi)$ variables.
The symplectic form in these variables can be directly read from
\(formaprereducida), and is given by
$$\Omega = -i d\omega^A\wedge d\pi_A + c.c.\(finalform)$$
The kernel of the symplectic form induced on the
constrained surface defined by the first-class constraint \(conspin)
precisely generates the phase shifts in $\omega$ and $\bar\pi$.
So if ${\bb T}$ (for twistor space \[Penrose]) denotes the four dimensional
complex vector space on which $\omega^A$ and $\bar\pi_{\dot A}$
are independent linear coordinates it follows that $M^{+}_{\alpha}$ is
nothing but the reduction of ${\bb T}$ with respect to the first-class
constraint \(conspin).

It is a standard result from the theory of coadjoint orbits
that the above phase space can be identified
with the coadjoint orbit of the Poincar\'e group associated with
massless particles with helicity $\alpha$ and future pointing momentum,
with $\Omega$ being the Kirillov-Kostant symplectic structure
associated with those orbits.

A completely analogous analysis can be carried out
for $M^+_{-\alpha}$ yielding a similar result up to the relative sign of
the helicity. The identification with the associated Poincar\'e orbits is,
of course, maintained.

Due to the relationship of these orbits with twistor space one can give an
alternative description of them in twistor variables as follows
\[PenroseII].
If $Z$ represents the pair of spinors $(\omega ,\bar\pi )$, then
one can take as twistor components
$$Z^{\gamma} =(\omega^0,\omega^1, \bar\pi_{\dot 0},\bar\pi_{\dot 1}).\()$$
If one defines the conjugate twistor $\bar Z$ by
$$\bar Z_{\gamma} =(\pi_0 ,\pi_1,\bar\omega^{\dot 0},\bar\omega^{\dot 1}),
\()$$
the constraint \(conspin) can be simply written as $Z^{\gamma}\bar Z_{\gamma}
=2\alpha$.
Finally the symplectic form can be written as
$$-i dZ^{\gamma}\wedge d\bar Z_{\gamma} + c.c.\()$$

With all of this in mind we will now pass to quantize the rigid particle model.

\section{Quantization}

Due to the identification of the reduced phase space of the rigid particle
model
with the coadjoint orbits of the Poincar\'e group for massless particles with
helicity $s$, the quantization of this system is an exercise that has already
been the object of study in standard textbooks. It will be certainly
out of the scope of this paper to give a full account of the
standard procedures,
and for that we will refer the reader to the excellent book of Woodhouse
on geometric quantization \[Woodhouse]. Anyhow, as the full machinery of
geometric quantization is not entirely necessary to understand the
quantization of such a simple system, we will attempt here to extract
from \[Woodhouse] the bare essentials.

The covariant quantization of the model will now be performed {\it \`a la}
Dirac, by imposing the first-class constraint \(conspin) on the physical
states.
First we start by choosing a polarization generated by $\partial /\partial
\omega$ and its complex conjugate, {\ie}, we will choose our wave functions to
be functions of $\pi$ and $\bar\pi$. In this representation it is
obvious that the operator associated with $\omega$ becomes
$$\omega^A\rightarrow -{\partial\ \over\partial\pi_A},
\quad
\bar\omega^{\dot A}\rightarrow {\partial\ \over\partial\bar\pi_{\dot A}},\()$$
while the operator associated with $\pi$ is simply given by multiplication
by $\pi$.

The constraint \(conspin) can be now directly written as
$$
\bar\pi_{\dot A} {\partial\ \over\partial\bar\pi_{\dot A}}-
\pi_A{\partial\ \over\partial\pi_A} =2\alpha.\(consquan)
$$
It is easy to show that this expression does not suffer from
ordering ambiguities provided that we choose the same ordering
for the two pairs $(\pi,\omega)$ and $(\bar\pi,\bar\omega)$,
since the right-hand-side of their respective commutators have
opposite signs and hence the ordering ambiguities cancel.

The quantization of $\alpha$ can now be understood in several ways. The more
geometrical one is associated with the integrality condition of the
symplectic potential
$$\theta = -i \omega^A d\pi_A + c.c.\()$$
in the constrained manifold. But it can also be understood in a more
standard physical way by showing the equivalence of the associated
Hilbert space with the solution space of massless wave equations of
arbitrary spin. Indeed,
for positive helicity, the wave functions $\varphi (\pi,\bar\pi )$
obeying the constraint \(consquan) can be mapped consistently into
the positive frequency solutions of the left-handed massless wave
equation \[Penrose]
$$\nabla^{A\dot A}\varphi_{AB...C} (\x )=0\(masswave)$$
by means of the Fourier transform
$$\varphi_{AB...C} (\x ) =
\left( {1\over 2\pi}\right)^{3\over 2} \int_{N^+} \varphi (\pi,\bar\pi)
\pi_A\pi_B...\pi_C\
{\rm e}^{-i \P\x} d\tau,\(sol)$$
where there are $2\alpha$ $\pi$'s, and the integration is over the
future light cone with $d\tau$ its natural Lorentz invariant measure
$$d\tau ={dP_1\wedge dP_2\wedge dP_3\over |P_0|}.\()$$
Because of the homogeneity properties
of $\varphi (\pi,\bar\pi )$ the integrand is well defined in the
constrained surface, {\ie}, it is impervious to transformations of the
form $\pi\rightarrow {\rm e}^{i\beta}\pi$ for $\beta\in{\bb R}$.

The right-handed solutions can be equally obtained by recalling the
natural correspondence between $M^+_{-\alpha}$ and $M^-_{\alpha}$.

A quantization in the twistor polarisation has been already pursued in simple
terms in \[PenroseII].
The interested reader can find there all the required
information, so we will avoid here any unnecessary repetition.

We would like to stress, as a final remark, that the conformal
invariance of these massless spin equations has a natural counterpart in the
particle model. It is evident from the definition of the action \(rigidaction)
that it only depends on the conformal class of the Minkowski metric.

\subsection{In the search for gauge invariance}

Although from the quantization of the Poincar\'e orbit for $\alpha=1$ one
obtains directly Maxwell equations in
spinor form, as a physicist, one is a little disappointed
by the fact that the gauge potential does not seem to come out from
the formalism.
As we will see below not only the gauge potential is naturally there,
but we will be able to interpret its associated $U(1)$ gauge transformations
as a direct consequence of the constraint structure of the model
arising from its $\W_3$ symmetry.

In order to see how this happens one should return to the (pre)symplectic
two-form in spinor coordinates \(omegaspin)
$$ -i\alpha {d\eta_A\wedge d\xi^A\over \xi^C\eta_C}
+i\alpha {\eta_A\xi^B d\eta_B\wedge d\xi^A\over (\xi^C\eta_C)^2}
+ c.c. \(omegaspindos)$$
The key observation is that we can greatly simplify its expression
after a suitable redefinition of constraints.
Indeed, notice that the two-form \(omegaspindos) reduces to
$$
-i d\eta_A\wedge d\xi^A + c.c.\(preomegaspin)
$$
on the submanifold defined by the first-class constraint
$\xi^A\eta_A=\alpha$. That one can impose consistently this
condition follows directly from the fact that $\xi$ and $\eta$
are defined only up to multiplication by an arbitrary complex
number, and  that $\xi^A\eta_A$ cannot be zero (otherwise
$\z\cdot\bar\z=0$).
It can also be easily checked that the kernel of
\(preomegaspin) on the constrained submanifold generates the remaining
freedom left in the spinors.
The constraint above reduces the arbitrariness in the spinors
to $\xi\rightarrow a\xi$ and $\eta\rightarrow (1/a)\eta$.
One can check now that exactly those gauge orbits are the ones generated
by the hamiltonian vector fields associated with the constraint and its
complex conjugate.

The way to recover the gauge potential should be by now clear: one should
quantize the phase space with coordinates $(\x, \P, \eta^A, \xi^A)$,
symplectic form
$$ d\x\wedge d\P -i d\eta_A\wedge d\xi^A + c.c.\()$$
and subject to the first-class constraints\fnote{\tiny Notice that due
to the complex character of $\psi_1$ and $\psi_2$ the counting
of the number of degrees of freedom yields the correct result.}
$$\psi_1 =\eta_A\xi^A-\alpha,\quad \psi_2 =P_{B\dot
B}\xi^B\bar\eta^{\dot B}, \quad{\rm and}\quad \psi_3=\P^2,$$
with Dirac's prescription for first-class constraints.

Explicitly, if one takes the natural polarization associated with the
symplectic potential
$$\theta =-\P d\x - i\eta_A d\xi^A - i\bar\xi_{\dot U}
d\bar\eta^{\dot U},\()$$
one has that under quantization
$$ \x\rightarrow -i{\partial\ \over\partial \P},
\quad \xi_A\rightarrow {\partial\ \over\partial\eta^A},
\quad {\rm and} \quad\bar\eta_{\dot A}\rightarrow{\partial\ \over\partial
\bar\xi^{\dot A}},\()$$
while $\P$, $\eta$ and $\bar\xi$ go to standard multiplication operators
acting on wave functions $A(\P,\eta,\bar\xi)$.

If we now impose the constraints as operator identities on the wave
functions we obtain
$$\eqnalign{\eta^A {\partial\ \over\partial\eta^A}\cdot A(\P,\eta,\bar\xi)=&
\alpha A(\P,\eta,\bar\xi),\()\cr
\bar\xi^{\dot A} {\partial\ \over\partial\bar\xi^{\dot A}}\cdot
A(\P,\eta,\bar\xi)=&
\alpha A(\P,\eta,\bar\xi),\()\cr
P_{A\dot A}\eta^A\bar\xi^{\dot A} A(\P,\eta,\bar\xi)=&0,\()\cr
P^{A\dot A}{\partial\ \over\partial\eta^A}{\partial\ \over\partial
\bar\xi^{\dot A}}\cdot A(\P,\eta,\bar\xi)=& 0,\()\cr
P^2 A(\P,\eta,\bar\xi)=& 0.\()\cr}$$
Notice that in this case the quantization of the constraints
$\psi_1$ and $\bar\psi_1$ suffers from ordering ambiguities and here
we have chosen to write all derivative operators on the right;
the only choice consistent with the covariantly reduced space
quantization of the previous section.
These ordering ambiguities however do not affect the first-class
character of the quantum constraint algebra.

The first two constraints for $\alpha=1$ simply tell us that
the wave function is of the form
$$A(\P,\eta,\bar\xi) = A_{B\dot B}(\P ) \eta^B\bar\xi^{\dot B}.\()$$
But notice from the third condition that $A_{B \dot B} (\P )$ is only
defined up to a term
of the form $P_{B\dot B}\varphi (\P )$,
and this is nothing but the standard gauge transformation of the vector
potential in momentum space. The remaining two constraints can be now seen
to impose the Lorenz gauge condition and the mass shell condition
respectively.

The above result can be given a clear geometrical interpretation. The
gauge invariance of the wave functions comes as a direct consequence of
the fact that the trajectories in the rigid particle model are not
physical. This agrees with the intuitive idea that gauge invariance
improves the renormalizabity properties of the associated quantum
theory by delocalizing the position of the photon. The rigid particle
model provide us with a precise mathematical description of this
intuitive physical idea.

The details of the generalization of the above procedure for
other integer values of $\alpha$ will be left
as an exercise for the interested reader. It is obvious from the previous
construction that only the particular form of the wave functions and their
corresponding gauge invariances will depend on the particular value
of $\alpha$, not so the quantization procedure sketched above.
We will finish this section by stating that, in particular,
for $\alpha =2$ one obtains linearized Einstein gravity in
terms of the (traceless) metric deviation from flat space-time in
the Einstein gauge. In this case the corresponding
gauge invariance is nothing but linearized general covariance.

\section{Final comments}

We hope to have convinced the reader that the rigid particle model and its
associated $\W_3$ symmetry play an important role in the physics of
massless particle models in four dimensions,
as well as in the geometry of gauge invariance.
It is also, in our opinion, quite remarkable that a purely ``bosonic"
particle model is suitable for the description of Weyl fermions in four
dimensional Minkowski space-time. Thus opening the door to the understanding
of four-dimensional Fermi-Bose transmutation in terms of this system
(for a somehow related approach see \[Wiegmann]\[Forte]).

It is natural to wonder if a similar approach can be developed for the
massive case. Unfortunately, the adding of an explicit mass term to the
rigid particle model leads to a reduced phase space without the adequate
dimensions \[Bacry]. It is therefore an open problem to find a geometrical
particle model which can be naturally associated with massive particles of
arbitrary spin.

It would be also interesting to investigate if more general geometrical
particle models can incorporate naturally, under quantization, a bigger
gauge invariance group than $U(1)$. Under the condition of locality and
invariance under conformal rescaling of the metric, a property that
should be preserved if one wishes to obtain conformally invariant field
theories, the most general four dimensional particle action
is of the form
$$S =\sum_{i=1}^3 \alpha_i \int_{\gamma} \kappa_i,\(parida)$$
where the $\kappa$'s correspond to the generalized curvature functions
associated with the path $\gamma$. The required phase space has
dimension $32$, although of course a plethora of constraints will
naturally arise from \(parida). The structure of the reduced phase
space for particular values of the parameters $\alpha_i$ can a priori
host non-abelian gauge invariance---a possibility that is under
current investigation.

\vskip 0.5truecm

\ack
We would like to thank J.M. Figueroa-O'Farrill, J. Gomis and J. Mas for
many useful discussions on the subject.  We are also grateful to the
Spanish Ministry of
Education, the British Council and the Human Capital and Mobility
program for financial support.
\vskip 0.5truecm

\refsout
\bye